\documentclass[10pt,journal,twocolumn]{IEEEtran}
\IEEEoverridecommandlockouts
\usepackage{algpseudocode}
\usepackage{algorithm}
\usepackage{bm}
\usepackage{amsfonts}
\usepackage{amssymb}
\usepackage{amscd}
\usepackage{graphics}
\usepackage[dvips]{graphicx}
\usepackage{epsfig}
\usepackage{subfigure}
\usepackage[cmex10]{amsmath}
\usepackage{array}
\usepackage{eqparbox}
\usepackage{flushend}
\usepackage{fancyhdr}
\usepackage{titling}
\usepackage{enumerate}
\usepackage{stackrel}

\DeclareMathAlphabet{\mathbsf}{OT1}{cmss}{bx}{n}
\DeclareMathAlphabet{\mathssf}{OT1}{cmss}{m}{sl}
\DeclareMathAlphabet{\mathcsf}{OT1}{cmss}{sbc}{n}

\newcommand{\ie}{{\em i.e.}}
\newcommand{\etc}{{\em etc}}
\newcommand{\eg}{{\em e.g.}}

\newcommand{\secref}[1]{Section~\ref{#1}}
\newcommand{\figref}[1]{Fig.~\ref{#1}}

\newcommand{\appref}[1]{Appendix~\ref{#1}}

\makeatletter
\def\blfootnote{\xdef\@thefnmark{}\@footnotetext}
\makeatother

\renewcommand{\baselinestretch}{1}
\newtheorem{theorem}{Theorem}[section]

\newtheorem{note}[theorem]{Note}

\newcommand{\qed}{\nobreak \ifvmode \relax \else
      \ifdim\lastskip<1.5em \hskip-\lastskip
      \hskip1.5em plus0em minus0.5em \fi \nobreak
      \vrule height0.75em width0.5em depth0.25em\fi}

\def\BibTeX{{\rm B\kern-.05em{\sc i\kern-.025em b}\kern-.08em
    t\kern-.1667em\lower.7ex\hbox{E}\kern-.125emX}}

\usepackage{newlfont}
\begin{document}
\title{A PMU Scheduling Scheme for Transmission of Synchrophasor Data in Electric Power Systems}
\author{\IEEEauthorblockN{K. G. Nagananda, Shalinee Kishore and Rick S. Blum}\thanks{K. G. Nagananda is with PES University, Bangalore 560085, INDIA, E-mail: \texttt{kgnagananda@pes.edu}. Shalinee Kishore and Rick S. Blum are with Lehigh University, Bethlehem, PA $18015$, U.S.A. E-mail: \texttt{\{skishore,rblum\}@lehigh.edu}}}

\pagenumbering{gobble}
\date{}
\setlength{\droptitle}{-0.5in}
\maketitle

\begin{abstract}
With the proposition to install a large number of phasor measurement units (PMUs) in the future power grid, it is essential to provide robust communications infrastructure for phasor data across the network. We make progress in this direction by devising a simple time division multiplexing scheme for transmitting phasor data from the PMUs to a central server: Time is divided into frames and the PMUs take turns to transmit to the control center within the time frame. The main contribution of this work is a scheduling policy based on which PMU transmissions are ordered during a time frame.

The scheduling scheme is independent of the approach taken to solve the PMU placement problem, and unlike strategies devised for conventional communications, it is intended for the power network since it is fully governed by the measure of electrical connectedness between buses in the grid. To quantify the performance of the scheduling scheme, we couple it with a fault detection algorithm used to detect changes in the susceptance parameters in the grid. Results demonstrate that scheduling the PMU transmissions leads to an improved performance of the fault detection scheme compared to PMUs transmitting at random.
\end{abstract}

\vspace{-0.15in}
\section{Introduction}\label{sec:introduction}
In order to provide near real-time wide area monitoring and control of power systems, synchrophasor data from PMUs are provided from across the power system by electric utilities. Typically, the PMUs are designed to record up to 30 - 60 measurements/second, and phasor data are transmitted to a centrally located wide area monitoring system (WAMS) server, where they are archived and recovered for several applications \cite{Phadke2008}. The high frequency of measurements by the PMUs and the applications involving synchrophasor data, together with the proposition to populate the future grid by a large number of PMUs, necessitates reliable and robust communications infrastructure within the power network \cite{Phadke2007}, \cite{Kirti2007}. In this paper, we make progress in this direction by devising a simple yet reliable method for transmission of phasor data from the PMUs to the WAMS server over dedicated direct communications links. We begin by summarizing the problem setup and the methodology developed to achieve the desired objective.

\begin{figure}[t]
\includegraphics[height=1.25in,width=3.5in]{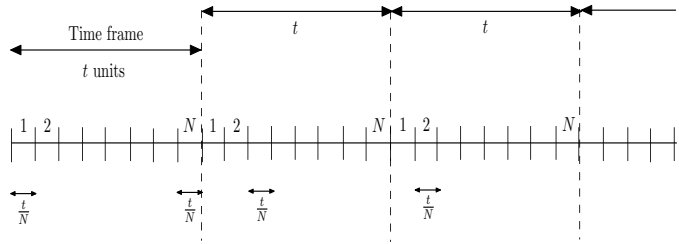}
\caption{Time division multiplexing of PMU transmission. }
\label{fig:pmu_trans1}
\end{figure}

Consider $N$ PMUs installed on the power network. Let time be divided into frames with the duration of each frame equal to $t$ units. A time frame is further divided into $N$ slots, each of duration $\frac{t}{N}$ time units. Within a time frame, the $N$ PMUs transmit phasor data to the WAMS server via dedicated channels of finite capacity. We illustrate this setup in \figref{fig:pmu_trans1}. In the communications theory literature, this is commonly referred to as \emph{time division multiplexing}, and the scheme incurs a delay of $(N-1)t/N$ time units per frame for each PMU. Given this setup, a fundamental question that arises is the following: What is the order in which these $N$ PMUs transmit to the WAMS server? In other words, what is the transmission schedule for the $N$ PMUs, so that the WAMS server can use the received data from the ordered set of PMUs to more quickly and more reliably determine changes in the system state. We investigate this question in this paper.

Scheduling policies intended for collecting data from PMUs should take into account the electrical properties of the power network being monitored. More precisely, the policy should be governed by the measure of electrical influence or connectedness between various network components. One way to measure the connectivity is to characterize the electrical coupling between buses in the network; the coupling can be obtained by computing the magnitude of the entries of the singular vectors obtained from the singular value decomposition (SVD) of the network matrices \cite{Wang2010}.

We answer the aforementioned question in two steps:
\begin{enumerate}[(1)]
    \item The first step is the classic PMU placement/seclection problem of obtaining the optimal number ($N$) of PMUs with the goal to have either complete or incomplete network observability either in the presence/absence of zero injection measurements. Here, we consider two cases:
        \begin{enumerate}[(a)]
        \item the topology-based PMU placement \cite{Nuqui2005} - \nocite{Gou2008}\nocite{Gou2008a}\nocite{Baldwin1993}\nocite{Milosevic2003}\nocite{Zhang2010}\nocite{Xu2004}\nocite{Azizi2012}
        \nocite{Kekatos2012}\nocite{Li2013}\nocite{Fesharaki2013}\cite{Anderson2014}, where the optimal number of $N$ PMUs is obtained from the node degree distribution of the grid; \label{case:topology} and
        \item the electrical structure-based approach to PMU placement, which was first adopted in \cite{Nagananda2014}. \label{case:electrical}
        \end{enumerate}
    \item Next, we devise the scheduling scheme for the $N$ PMUs. If $B$ is the number of buses in the network, we construct the $B\times B$ bus admittance and resistance distance matrices (see \cite[Section III]{Cotilla-Sanchez2012}). For case (\ref{case:topology}) stated above, we pick the $N \times N$ sub-matrix of the $B\times B$ bus admittance matrix with the rows and columns corresponding to bus locations where the PMUs are installed. We perform SVD of this sub-matrix; the absolute values of elements of the resulting $N$ singular vectors are central to devising the ordering strategy for PMU transmissions. For  (\ref{case:electrical}), this procedure is repeated on the resistance distance matrix.
\end{enumerate}

In order to quantify the performance of the scheduling scheme, we couple it with a fault detection algorithm in which changes in the bus susceptance parameters are detected. The detection problem is formulated using a linear errors-in-variables model, and a generalized likelihood ratio test (GLRT) based on the total least squares (TLS) methodology is presented. The performance of TLS-GLRT is analyzed with and without the proposed scheduling policy. Results demonstrate that scheduling PMU transmissions leads to an improvement in the probability of fault detection.

Some advantages of the proposed scheduling policy are:
\begin{enumerate}[1)]
\item The topology-based approach to PMU placement (case (\ref{case:topology})) incorporates less known information, since it neglects the sensitivity between power injections and nodal phase angle differences, while case (\ref{case:electrical}) is based on the complex networks perspective of the power grid, and was shown to provide a more comprehensive characterization of the electrical influence between network components (see \cite{Cotilla-Sanchez2012}). However, the general framework of the scheduling policy derived in this work remains unchanged for both cases.
\item In practice, there is a layer of phasor data concentrators (PDCs) between the PMUs and the WAMS server. Our scheduling policy  is unaffected by the presence of PDCs, since it is devised from a transmitter-centric viewpoint.
\end{enumerate}
\vspace{-0.2in}
\subsection{Scheduling in the power grid}
There are different types of scheduling in power networks which have been widely examined in the literature. For instance, there are architectures for power scheduling, algorithms for traffic ({\eg}, multimedia data) scheduling on the grid, user-access scheduling procedures for smart power appliances, {\etc}. In this following we point to a few references, where each paper concerns a specific type of scheduling on the grid. In the interest of space we restrict ourselves to four references, with due credit to other valuable contributions.

In \cite{Zhou2012}, the authors proposed a power scheduling scheme for the smart grid from the perspectives of architecture, strategy and methodology based on the quality of experience (QoE); the QoE metric quantifies the customers' degree of satisfaction. A novel approach to QoE modeling was proposed, and an automatic proactive in-service strategy was employed to estimate the end user's QoE. In \cite{He2013}, a multi-time scheduling scheme, in the framework of Markov decision processes, was proposed for two classes of energy users, namely, traditional and opportunistic energy users. The reliability of the power system operation was analyzed under supply uncertainty as a result of variable and non-stationary wind generation, demand uncertainty owing to the stochastic behavior of a large number of opportunistic users and the coupling between sequential decisions across multiple timescales.

In \cite{Huang2013}, the advantages provided by wireless multimedia sensor networks in conjunction with the benefits of cognitive radio technology were exploited to devise a priority-based scheduling scheme for smart grid traffic. The traffic types included control commands, multimedia sensing data and meter readings. A joint access and scheduling approach for in-home appliances (both schedulable and critical) was devised in \cite{Chen2013} to coordinate the power usage to keep the total energy demand for the home below a target value. Uncertainties in the variations of electricity prices and distributed wind power were incorporated into the scheduling scheme to optimize the performance of the energy management controller.

The scheduling scheme devised in this paper is different from the above mentioned works in that our scheme is aimed at ordering the transmission of PMUs for transfer of phasor data to the WAMS server. Our work is concerned with improving the communications efficiency of the set of PMUs installed on the grid to quickly and reliably detect changes in the system state; this paper does not deal with the power and/or traffic scheduling that have been addressed in the aforementioned references. To the best of our knowledge, this is the first instance where a scheduling scheme for PMU transmissions has been reported in the literature.

The remainder of this paper is organized as follows. In \secref{sec:pmu_placement}, we review the PMU placement problem, employing the topology- and electrical structure-based approaches. The PMU scheduling scheme is developed in \secref{sec:schedulingPMUs}. In \secref{sec:fault_detection}, we present the fault detection framework. \secref{sec:results} includes simulation results and related discussion. \secref{sec:conclusion} concludes the paper. The advantages of using the electrical structure of the grid over its topological structure, and the construction of the resistance distance matrix are relegated to \appref{app:electrical_structure}.
\section{The PMU placement problem}\label{sec:pmu_placement}
In this section, we revisit the PMU placement problem from two different perspectives: (a) topology-based approach and (b) electrical structure-based approach.

In the general setting, for a power network with $B$ buses and $K$ branches, and for complete network observability without zero injection measurements\footnote{Zero injection measurements are present when the power system has nodes without generation or load.}, the PMU placement problem is formulated as an integer linear program as follows \cite{Gou2008}, \cite{Gou2008a}:
\begin{eqnarray}\label{eq:pmuplacement}
\nonumber \min \sum_{i=1}^{B}d_i\\
\text{such that}~ \bm{C}\bm{d} &\geq& \bm{1},
\end{eqnarray}
with
\begin{eqnarray}\label{eq:binarydecision_vector}
d_{i} = \begin{cases}
1, ~\text{if a PMU is installed at bus}~i,\\
0, ~\text{otherwise}.
\end{cases}
\end{eqnarray}
$\bm{C}$ is the $B\times B$ binary connectivity matrix of the grid, $\bm{1}$ denotes a $B\times 1$ vector of 1's. The solution to \eqref{eq:binarydecision_vector} gives the optimal number $(N)$ of PMUs to be installed on the grid.
\begin{enumerate}[(1)]
    \item For PMU placement based on the topology of the grid, the entries of the bus admittance matrix are transformed into binary form and used in the problem setup \eqref{eq:pmuplacement}. In this case, $\bm{C}$ is given by
    \begin{eqnarray}\label{eq:topological_A}
    \bm{C}:
    \begin{cases}
    c_{ij} = 1, ~\text{if}~i=j,\\
    c_{ij} = 1, ~\text{if}~i~\text{and}~j~\text{are connected},\\
    c_{ij} = 0, ~\text{if}~i~\text{and}~j~\text{are not connected}.
    \end{cases}
    \end{eqnarray}
    The entries $c_{ij}$ of $\bm{C}$ characterize the electrical connections between network buses $i$ and $j$.

    \item For the electrical structure-based PMU placement, matrix $\bm{C}$ is derived as shown in \appref        {app:electrical_structure} (see \eqref{eq:adjcency_matrix}), and this will be used in the formulation \eqref{eq:pmuplacement}. The entries $c_{ij}$ of $\bm{C}$ are obtained taking into account the sensitivity between power injections and nodal phase angles differences between various buses in the grid.
\end{enumerate}
\vspace{-0.1in}
\section{Scheduling policy for PMU transmission}\label{sec:schedulingPMUs}\vspace{-0.05in}
In this section, we present the scheduling scheme so that the optimal $N$ PMUs have a predefined order to transmit phasor data to the WAMS server. In this work, we only consider complete network observability, and without zero injection measurements. In the following, $b = 1,\dots,B$ is the bus number index, while $n = 1,\dots,N$ is the index of the optimal number of PMUs. We devise the algorithm for the $N$ PMUs obtained from the topological structure-based placement (case (\ref{case:topology}) in the previous section). The same scheme is readily applicable for the case where PMU placement is solved by employing the electrical structure-based approach (case (\ref{case:electrical})).

The following is a step-by-step procedure for the proposed PMU scheduling policy:
\begin{enumerate}[1.]
    \item Obtain the optimal number $(N)$ of PMUs by solving the PMU placement problem \eqref{eq:pmuplacement}. \label{step1}
    \item In the $B\times B$ bus admittance matrix, pick those rows and columns which correspond to the bus numbers where PMUs are installed. We, therefore, have an $N \times N$ sub-matrix. \label{step2}
    \item Perform the SVD of the $N \times N$ sub-matrix to obtain the singular values and singular vectors. The $N \times 1$ left and right singular vectors are denoted $\bm{u}_n$ and $\bm{v}_n$, respectively, while the singular values are denoted $\sigma_n$. \label{step3}
    \item Compute the magnitude of the elements of the vectors $\sigma_n\bm{u}_{n}$. Note that, the index of each entry of the vector $\sigma_n\bm{u}_{n}$ corresponds to a bus location where a PMU is installed. 
    \item In the vector $\sigma_1\bm{u}_{1}$, {\ie}, the first column of the $N \times N$ sub-matrix, the PMU placed on the entry with the highest magnitude transmits first. Note that, $\bm{u}_{1}$ is the eigenvector corresponding to the largest eigenvalue.
        \label{transmit_first}
    \item The procedure in Step \ref{transmit_first} is repeated for the remaining vectors $\sigma_n\bm{u}_{n}$, $n = 2,\dots,N$, where the $\bm{u}_{n}$s are picked in the decreasing order of the corresponding eigenvalues.  \label{conflict}
\end{enumerate}

In Step \ref{conflict}, there is a possibility of conflict, which is explained via an example. Consider two vectors $\sigma_1\bm{u}_{1}$ and $\sigma_3\bm{u}_{3}$ used to schedule the transmission of PMUs during the first and third time slots, respectively. Suppose the entry having the largest magnitude in vector $\sigma_1\bm{u}_{1}$ is the same as the entry having the largest magnitude in vector $\sigma_3\bm{u}_{3}$. Then, the scheduling scheme picks the same PMU for both (first and third) time slots. To resolve this conflict, we propose the following modification to the scheduling scheme: for the third time slot, pick the entry in the vector $\sigma_3\bm{u}_{3}$ having the \emph{second largest magnitude}. If this entry is not the same as the one in vector $\sigma_2\bm{u}_{2}$ (used to schedule a PMU transmission for the second time slot), then the PMU placed on that entry is scheduled to transmit in the third time slot. This simple procedure is implemented for all the vectors $\sigma_n\bm{u}_{n}$. Note that, the priority given to PMU transmissions is solely based on the electrical connectedness of buses in the network, making it different from scheduling schemes devised for typical communications networks. In the next subsection, we explain the scheduling scheme via an illustration.
\vspace{-0.15in}
\subsection{An illustration}\label{subsec:illustration}
\begin{figure}[h]
\centering
  \includegraphics[height=2.5in,width=3.25in]{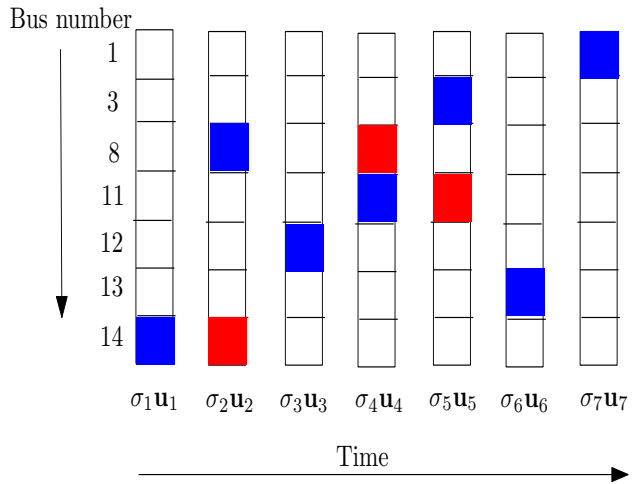}
  \caption{Scheduling PMUs for transmission for the IEEE 14-bus network.}
  \label{fig:pmu_schedule_14bus}
\end{figure}
We consider the IEEE 14-bus system to illustrate the scheduling scheme. We obtain the optimal number of PMUs to be placed on the network employing the electrical structure-based approach to PMU placement. For sake of brevity, we consider a single time frame, and implement the following steps to schedule the transmission of PMUs:
\begin{enumerate}[(1)]
\item For the 14-bus network, solving \eqref{eq:pmuplacement} yields an optimum of $N$ = 7 PMUs to be placed on buses numbered 1, 3, 8, 11, 12, 13 and 14 for complete network observability without zero injection measurements.

\item In the 14 $\times$ 14 resistance distance matrix, pick the rows and columns numbered 1, 3, 8, 11, 12, 13 and 14, thereby yielding a 7 $\times$ 7 sub-matrix.

\item Perform the SVD of the 7 $\times$ 7 sub-matrix to obtain the 7 $\times$ 1 right and left singular vectors $\bm{u}_n$ and $\bm{v}_n$, respectively, and the singular values $\sigma_n$, $n$ = 1, \dots, 7. Compute the magnitude of the elements of the vectors $\sigma_n\bm{u}_{n}$. The index of each entry of the vector $\sigma_n\bm{u}_{n}$ corresponds to a bus location where a PMU is installed. The seven vectors $\sigma_n\bm{u}_{n}$, $n$ = 1, \dots, 7 are depicted in \figref{fig:pmu_schedule_14bus}, where a column denotes a vector, while a box in each column denotes an entry of the vector. The number of boxes in each column equals the number $N$ of PMUs installed on the bus system.

\item In the vector $\sigma_1\bm{u}_{1}$ (the first column in \figref{fig:pmu_schedule_14bus}), the entry having the largest magnitude appears in the last row - marked in blue. Thus, the PMU placed on bus numbered 14 is scheduled to transmit in the first time slot.
\item In the vector $\sigma_2\bm{u}_{2}$ (the second column), the entry having the largest magnitude appears in the last row, similar to that in the vector $\sigma_1\bm{u}_{1}$, again allocating the PMU placed on bus numbered 14 to transmit in the second time slot. However, as described in the scheduling policy, this conflict is resolved by scheduling the PMU placed on the bus numbered 8, which has the second largest magnitude in the vector $\sigma_2\bm{u}_{2}$, to transmit in the second time slot.

\item Continuing in this fashion, and employing the conflict-resolution strategy, the PMUs installed on buses 14, 8, 12, 11, 3, 13 and 1 are scheduled to transmit in time slots numbered 1, 2, 3, 4, 5, 6 and 7, respectively.
\end{enumerate}
As mentioned in \secref{sec:introduction}, the scheduling policy is independent of (i) the approach (topology-based or electrical structure-based) taken to address the PMU placement problem, and (ii) the dynamic nature of the power system states. In the next section,  we analyze the impact of the scheduling scheme when it is incorporated into a change detection framework within the power network.
\vspace{-0.1in}
\section{A fault detection procedure}\label{sec:fault_detection}
The power system under test is modeled using the direct current (DC) power flow model with a linear relation between the active power flow on a transmission line and the difference of the voltage angles on the two corresponding buses \cite{Grainger2003}:
\begin{eqnarray}
\text{power flow} = \text{susceptance} \times \text{voltage angle difference}. \label{eq:dc_model}
\end{eqnarray}
Given the noisy measurements of voltage angle differences and power flows across various lines in the power network, we seek to detect changes in the susceptance parameters, {\ie}, whether the susceptance are equal to some nominal known values, or have changed. We pose the change detection as a hypothesis testing problem employing a linear error-in-variables (EIV) model, which allows for noise in both sides of the linear relationship \eqref{eq:dc_model}. The standard approach to parameter estimation
in such problems is known as total least squares (TLS) \cite{Huffel2002}, while hypothesis testing in EIV models have been addressed by deriving the generalized likelihood ratio tests (GLRT) \cite{Huang2001}. The TLS-GLRT to detect changes in the susceptance parameters of a grid was first proposed in \cite{Wei2012}.

The PMUs are typically used to obtain the noisy measurements of voltage angle differences and power flows across the network. In this section, we analyze the performance of TLS-GLRT to detect changes in the susceptance parameters when these PMUs are constrained to follow the scheduling scheme to transmit the noisy measurements to a fusion center, where the detection algorithm is implemented. \secref{subsec:tlsglrt_problem} comprises the problem formulation, while the TLS-GLRT solution is presented in \secref{subsec:tlsglrt_test}.
\vspace{-0.15in}
\subsection{Problem formulation}\label{subsec:tlsglrt_problem}\vspace{-0.05in}
Let us consider a power system with $B$ buses and $K$ branches in the network, which can be modeled as an undirected graph $\mathcal{G} = (\mathcal{B}, \mathcal{K})$ with $\mathcal{B} \triangleq \{1,\dots,B\}$ and $\mathcal{K} \triangleq \{(i_1, j_1),\dots,(i_K, j_K)\}$ denoting the sets of buses and branches, respectively. The $B \times 1$ vector $\bm{\theta}(t)$ and the $B \times B$ skew-symmetric matrix $\bm{Y}(t)$ models the voltage angles and the active power flow between the buses at time slot $t$, respectively. The $(i, j)^{\text{th}}$ entry of the matrix $\bm{S}$ describes the susceptance between buses $i$ and $j$: $\bm{S}_{ij} = \bm{S}_{ji}$ if $(i, j) \in \mathcal{K}$ and $\bm{S}_{ij} = 0$ otherwise. From \eqref{eq:dc_model}, we have
\begin{eqnarray}
\bm{Y}_{ij}(t) = \bm{S}_{ij}\left(\theta_i(t) - \theta_j(t)\right); t = 1,\dots,T. \label{eq:dc_powerflow}
\end{eqnarray}
Let $\bm{s}$ be a $K \times 1$ vector with elements $\bm{S}_{ij}$ and let $K \times 1$ vector $\bm{z}(t)$ be defined to collect $\bm{Y}_{ij}(t)$, for $(i, j) \in \mathcal{K}$ and $i > j$. Thus, \eqref{eq:dc_powerflow} can be written as
\begin{eqnarray}
\bm{z}(t) = \text{diag}(\bm{s})\bm{D}\bm{\theta}(t), \label{eq:dc_alternative}
\end{eqnarray}
where the $K \times B$ matrix $\bm{D}$ is defined as follows: for the $k^{\text{th}}$ branch $(i_k, j_k) \in \mathcal{K}$ and $i_k > j_k$, $\bm{D}_{k, i_k} = 1$ and $\bm{D}_{k, j_k} = -1$. The other elements in the $k^{\text{th}}$ row of $\bm{D}$ are zero. In practice, the noisy power flow and voltage angle measurements are given by
\begin{eqnarray}
\tilde{\bm{z}}(t) &=& \bm{z}(t) + \bm{w}_{z}(t), \label{eq:noisy_powerflow} \\
\tilde{\bm{\theta}}(t) &=& \bm{\theta}(t) + \bm{w}_{\theta}(t), \label{eq:noisy_voltageangle}
\end{eqnarray}
where the noise processes are given by
\begin{eqnarray}
\bm{w}_{z}(t) &\sim& \mathcal{N}(\mathbf{0}, \sigma_{z}^2\mathbf{I}), \\
\bm{w}_{\theta}(t) &\sim& \mathcal{N}(\mathbf{0}, \sigma_{\theta}^2\mathbf{I}).
\end{eqnarray}
In matrix notation, for $T$ time instants, we have
\begin{eqnarray}
\bm{Z} &=& \text{diag}(\bm{s})\bm{D}\bm{\Theta},\\
\tilde{\bm{Z}} &=& \bm{Z} + \bm{W}_z,\\
\tilde{\bm{\Theta}} &=& \bm{\Theta} + \bm{W}_{\theta}, \label{eq:matrix_notation}
\end{eqnarray}
where $\bm{Z}$ and $\bm{W}_z$ are $K \times T$ matrices used to collect $T$ samples of $\bm{z}(t)$ and $\bm{w}_z(t)$, respectively, while $\bm{\Theta}$ and $\bm{W}_{\theta}$ are of dimension $B \times T$ used to collect $T$ samples of $\bm{\theta}(t)$ and $\bm{w}_{\theta}(t)$, respectively.

The problem is to detect changes in the susceptance vector $\bm{s}$ based on the noisy observations $\tilde{\bm{Z}}$ and $\tilde{\bm{\Theta}}$. Towards this end, we assume knowledge of a vector $\bm{s}_0$ corresponding to the nominal behavior of the grid and test the following hypotheses:
\begin{eqnarray}
\begin{cases}
H_0: \bm{s} = \bm{s}_0\\
H_1: \bm{s} \neq \bm{s}_0.
\end{cases}\label{eq:hypothesis_test}
\end{eqnarray}
Under both hypotheses, $\bm{Z}$ and $\bm{\Theta}$ are unknown and have to be estimated.
\vspace{-0.15in}
\subsection{TLS-GLRT solution}\label{subsec:tlsglrt_test}
The TLS-GLRT solution to the aforementioned hypothesis testing problem assumes $\bm{\Theta}$, and therefore $\bm{Z}$, are deterministic unknown vectors. The maximum likelihood (ML) estimation of $\bm{s}$, $\bm{\Theta}$ and $\bm{Z}$ is therefore known as TLS \cite{Huffel2002}. The TLS-GLRT is given by
\begin{eqnarray}
    \text{t}_{\text{TLS}} = \log \frac{\max_{\bm{s},\bm{\Theta}}\prod_{t=1}^{T}p\left(\tilde{\bm{z}}(t), \tilde{\bm{\theta}}(t); \bm{s}, \bm{\theta}(t)\right)}{\max_{\bm{\Theta}}\prod_{t=1}^{T}p\left(\tilde{\bm{z}}(t), \tilde{\bm{\theta}}(t); \bm{s}_0, \bm{\theta}(t)\right)} \stackrel[H_0]{H_1}{\gtrless} \rho, \label{eq:tls_glrt}
\end{eqnarray}
where $\rho$ is a fixed threshold. We choose $H_0$ if the statistic is smaller than $\rho$, and H1 otherwise. The joint distribution of the observations is
\begin{eqnarray}
    p\left(\tilde{\bm{z}}(t), \tilde{\bm{\theta}}(t); \bm{s}, \bm{\theta}(t)\right)\!\! &=&\!\! p(\tilde{\bm{z}}(t); \bm{s}, \bm{\theta}(t))p(\tilde{\bm{\theta}}(t);\bm{\theta}(t)),\\
    p(\tilde{\bm{z}}(t); \bm{s}, \bm{\theta}(t)) &\sim& \mathcal{N}\left(\text{diag}(\bm{s})\bm{D}\bm{\theta}(t), \sigma_z^2\bm{I}\right),\\
    p(\tilde{\bm{\theta}}(t);\bm{\theta}(t)) &\sim& \mathcal{N}\left(\bm{\theta}(t), \sigma_{\theta}^2\bm{I}\right).
\end{eqnarray}

In simplified form, the TLS-GLRT is given by
\begin{eqnarray}
\nonumber\text{t}_{\text{TLS}} &=& \frac{1}{2}\text{Tr}\left\{\bm{A}^{\mathrm{T}}(\bm{s}_0)\bm{H}^{-1}(\bm{s}_0)\bm{A}(\bm{s}_0) \right\} \\ &&
 - \frac{1}{2}\min_{\bm{s}}\text{Tr}\left\{\bm{A}^{\mathrm{T}}(\bm{s})\bm{H}^{-1}(\bm{s})\bm{A}(\bm{s}) \right\} \stackrel[H_0]{H_1}{\gtrless} \rho, \label{eq:tls_glrt_simplified}\\
\bm{A}(\bm{s}) &=& \tilde{\bm{Z}} - \text{diag}\{\bm{s}\}\bm{D}\tilde{\bm{\Theta}},\\
\bm{H}(\bm{s}) &=& \sigma_{z}^2\mathbf{I} + \sigma_{\theta}^2\text{diag}\{\bm{s}\}\bm{D}\bm{D}^{\mathrm{T}}\text{diag}\{\bm{s}\}
\end{eqnarray}

The threshold $\rho$ is chosen as follows: given enough samples, the asymptotic performance of TLS-GLRT under $H_0$ is:
\begin{eqnarray}
2 \text{t}_{\text{TLS}} \sim \mathcal{X}^2_K, \label{eq:chisquared}
\end{eqnarray}
where $\mathcal{X}^2_K$ is a Chi-squared random variable with $K$ degrees of freedom, and $K$ is the dimension of $\bm{s}$ \cite{Kay2001}. This result is independent of the specific value of the unknown $\bm{\Theta}$. A reasonable approach to choosing the threshold for a given
false alarm rate $\alpha$ is
\begin{eqnarray}
\rho_{\alpha} = \frac{1}{2}F^{-1}_{\mathcal{X}^2_K}(\alpha), \label{eq:false_alarm}
\end{eqnarray}
where $F^{-1}_{\mathcal{X}^2_K}(.)$ is the inverse cumulative distributive function of the Chi-squared distribution with $K$ degrees of freedom.

Some comments on \eqref{eq:false_alarm} are in order. The physical meaning of \eqref{eq:false_alarm} is that if $T$ is large and the assumptions in \eqref{eq:dc_powerflow} - \eqref{eq:matrix_notation} hold, then the test statistic $2\text{t}_{\text{TLS}}$ in \eqref{eq:chisquared} can be shown to be a Chi-squared random variable with $K$ degrees of freedom.  Even for moderately large $T$, the approximation is often quite accurate. A Chi-squared random variable has an inverse cumulative distributive function $F^{-1}_{\mathcal{X}^2_K}(.)$ that is a common and extensively tabulated function available in many software packages and whose values are tabulated in books \cite[Chapter 2.2]{Kay2001}. Numerous efficient algorithms are available to compute this function, however, the computational complexity is not really an issue. In practice, one can chose a set of desirable false alarm probabilities, $\alpha$ in \eqref{eq:false_alarm}, and the corresponding thresholds, $\rho_{\alpha}$ in \eqref{eq:false_alarm}, can be computed off-line and stored in a look-up table. Then the system can choose any of these false alarm probabilities and this will determine the threshold to employ. Fixing the false alarm probability is accepted practice in hypothesis testing \cite{Kay2001}. Given the fixed false alarm probability, the test in \eqref{eq:tls_glrt_simplified} is chosen to optimize the probability of detection ($P_d$) as described in \cite{Wei2012}. Note that in  detection theory literature, the probability of false alarm is the probability of incorrectly choosing $H_1$ when $H_0$ is actually true, while the probability of deciding on hypothesis $H_1$ when $H_1$ is indeed true is referred to as the probability of detection \cite[Chapter 3]{Kay2001}.
\vspace{-0.1in}
\section{Simulation results and discussion}\label{sec:results}
In this section, we present simulation results to demonstrate the performance improvement of the TLS-GLRT fault detection scheme when the transmission of PMUs follow the scheduling policy described in \secref{sec:schedulingPMUs} compared to its performance when the PMUs transmit in a round-robin fashion without a predefined order. Results also enable us to compare the performance of the detection scheme offered by scheduling the PMU transmissions, when the PMUs are placed employing the topology- and electrical structure-based approaches. We first describe the experimental setup, followed by simulation results and related discussion.
\vspace{-0.1in}
\subsection{Experimental setup}\label{subsec:setup}
We consider the IEEE 14-bus system with the number of buses $B$ = 14 and the number of branches $K$ = 20. The noise variances are given by $\sigma_{z}^2$ = $\sigma_{\theta}^2$ = 0.01. We perform 10$^4$ Monte Carlo simulations. At each realization, the elements of $\bm{\Theta}$ are generated independently as standard normal random variables. The active power flow $\bm{Y}$ between the buses were obtained using the power flow algorithm in MATPOWER \cite{Zimmerman2011}; the susceptance vector $\bm{s}$ was then computed using $\bm{Y}$. Under $H_0$, we use the susceptance parameters given by the test profile. Under $H_1$, we apply a change of -2\% to every element of $\bm{s}$. Here, -2\% essentially means that every element of $\bm{s}$ is made smaller by a factor of 2\%; this number can be arbitrarily chosen without loss of generality.

Simulations are conducted for one frame lasting 20 time units. This corresponds to $T = 20$ in \eqref{eq:dc_powerflow}. Furthermore, each frame is subdivided into $N$ slots, where $N$ is the solution to the PMU placement problem. The setup comprises solving the hypothesis testing problem \eqref{eq:hypothesis_test} after each PMU transmission. Note that, we need phasor data from all the PMUs to form the test statistic. In light of the previous statement, we do not induce a change in the susceptance vector at the first time instant; however, for subsequent time instants, every element of the susceptance vector $\bm{s}$ will be changed by -2\%. After each transmission, we use the phasor data that we previously had from each PMU to form the test statistic.

To analyze the performance of the fault detection scheme presented in \secref{sec:fault_detection}, we compute the variation of probability of detection $(P_d)$ of change in the susceptance parameter versus time, using a fixed value of false alarm rate ($\alpha$) uniformly picked between [0, 0.2] for which the corresponding threshold is calculated using \eqref{eq:false_alarm}. We then pick a new value of $\alpha \in$ [0, 0.2] and repeat the calculations to obtain the corresponding variation of $P_d$ with time. This procedure is repeated for different values of $\alpha \in$ [0, 0.2], resulting in a set of $P_d$s versus time. Lastly, we plot the average (over the number of $\alpha$s) $P_d$ versus time for two scenarios: (a) when the PMUs transmit in a random manner without a prescribed scheduling policy and (b) when the PMUs transmissions are allowed to follow the scheduling scheme. For both scenarios, we consider the separate cases of the PMUs being placed employing the topology- and electrical structure-based approaches.
\vspace{-0.125in}
\subsection{Simulation results}\label{subsec:simulation_results}
We first present the results for the PMU placement problem for the 14-bus system. For the topology-based approach to the PMU placement problem, the connectivity matrix is given by \eqref{eq:topological_A}. Solving \eqref{eq:pmuplacement}, we obtain 4 as the optimum number of PMUs to be installed on buses 2, 6, 7 and 9. For the electrical structure-based approach to the PMU placement problem, the connectivity matrix is given by \eqref{eq:adjcency_matrix} (see \appref{app:electrical_structure}). Solving \eqref{eq:pmuplacement}, we obtain 7 as the optimum number of PMUs to be installed on buses 1, 3, 8, 11, 12, 13 and 14. For both topology- and electrical structure-based approaches, the objective is to achieve complete network observability.

Following the scheduling policy described in \secref{sec:schedulingPMUs}, for the topology-based approach (where an optimal number of 4 PMUs are placed on buses numbered 2, 6, 7 and 9), the transmission schedule is as follows: the PMU placed on bus 7 transmits first, followed by those placed on buses 2, 6 and lastly 9, in that order.

Similarly, for the electrical structure-based approach (where an optimal number of 7 PMUs are placed on buses numbered 1, 3, 8, 11, 12, 13 and 14), the transmission scheduled is the same as presented in \secref{subsec:illustration}, {\ie}, the PMUs installed on buses 14, 8, 12, 11, 3, 13 and 1 transmit in time slots numbered 1, 2, 3, 4, 5, 6 and 7, respectively.

\begin{figure}[t]
\includegraphics[height=2.5in,width=3.5in]{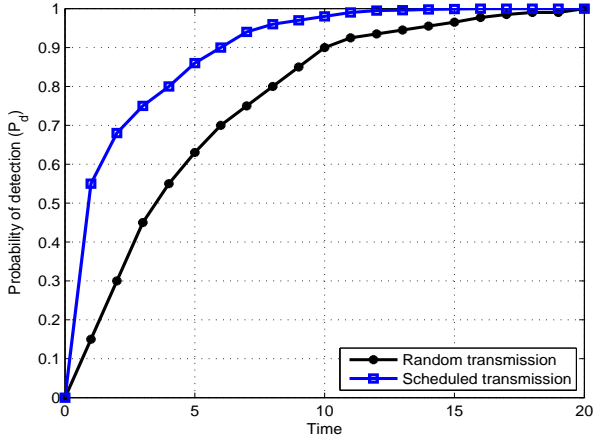}
\caption{Probability of detection versus time, for one frame lasting 20 time units, when the PMUs are placed employing the electrical structure-based approach.}
\label{fig:Pd_time_elec_14bus}
\end{figure}
\begin{figure}[t]
\includegraphics[height=2.5in,width=3.5in]{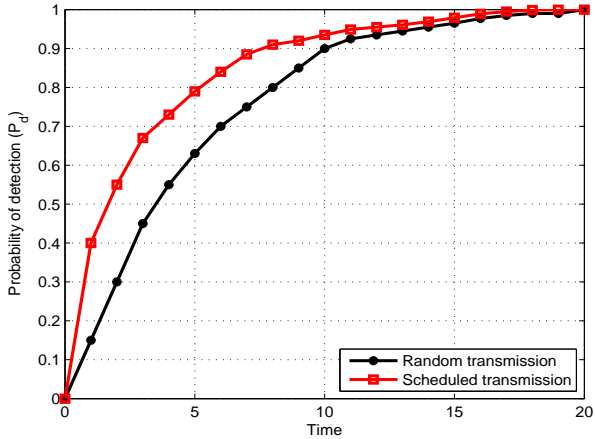}
\caption{Probability of detection versus time, for one frame lasting 20 time units, when the PMUs are placed employing the topology-based approach.}
\label{fig:Pd_time_topo_14bus}
\end{figure}
The plots of $P_d$ versus time for one frame lasting 20 time units when the PMUs are placed employing the electrical structure- and topology-based approaches are shown in \figref{fig:Pd_time_elec_14bus} and \figref{fig:Pd_time_topo_14bus}, respectively. In each figure, we also plot $P_d$ versus time when the PMU-transmissions are random, without a predefined order/schedule. As seen from the plots, scheduling the PMU-transmissions results in better detection performance compared to PMUs transmitting at random. Note that, at the end of the time frame, after the vector of phasor data from all the PMUs are updated, the $P_d$ for both scheduled and random transmissions become the same since we have phasor data from all the PMUs in both cases.

For the topology-based approach to PMU placement, we have installed 4 PMUs on the grid, while for the electrical structure-based approach, there are 7 PMUs. To compare the scheduling performance between the topology- and electrical structure-based PMU placements, we modify the experimental setup as follows. We let the same number of PMUs transmit for both the topology-based and electrical structure-based approaches to analyze the detection performance. Specifically, we allow only 4 PMUs transmit their phasor data to the control center for both PMU placement approaches before applying the scheduling scheme. More precisely, for the topology-based approach we allow PMUs on buses numbered 2, 6, 7 and 9 to transmit as before. For the electrical structure-based approach, we now allow only 4 PMUs on buses numbered 8, 11, 12 and 14 transmit their phasor data. This choice was based on the magnitudes of the entries in the vectors $\sigma_n\bm{u}_{n}$, $n=1,\dots,4$, {\ie}, the first four column vectors in the 7 $\times$ 7 sub-matrix described in \secref{subsec:illustration}.

\begin{figure}[t]
\includegraphics[height=2.5in,width=3.5in]{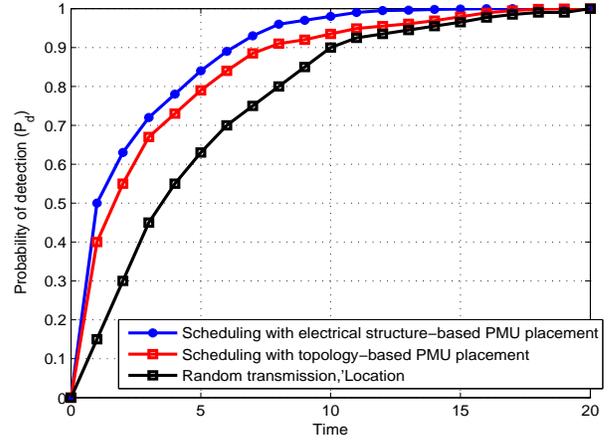}
\caption{Probability of detection versus time, for one frame lasting 20 time units, to compare the topology- and electrical structure-based approaches to PMU placement.}
\label{fig:Pd_time_elec_topo}
\end{figure}
The scheduling policy is now applied to the above setup. Similar to previous experiments, for the topology-based approach the transmission schedule is as follows: the PMU placed on bus 7 transmits first, followed by those placed on buses 2, 6 and lastly 9, in that order. For the electrical structure-based approach, the PMUs installed on buses 14, 8, 12 and 11 transmit, in that order. For both approaches, each frame (lasting 20 time units) is divided into 4 slots. The plot of $P_d$ versus time is shown in \figref{fig:Pd_time_elec_topo}, where we see that the electrical structure-based approach has a better detection performance compared to the topology-based approach.

\begin{figure}[t]
\includegraphics[height=2.5in,width=3.5in]{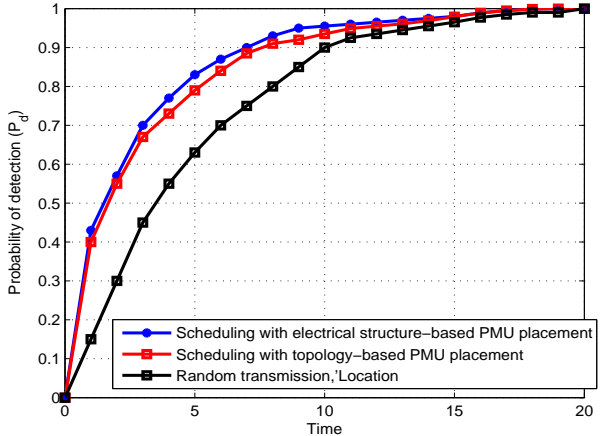}
\caption{Probability of detection versus time, for one frame lasting 20 time units, with lesser number of PMUs for the electrical structure-based approach compared to the topology-based approach.}
\label{fig:Pd_time_elecless_topo}
\end{figure}

Finally, we conduct an experiment in which we allow lesser number of PMUs for the electrical structure-based approach to transmit compared to the topology-based approach. Specifically, we allow only 3 PMUS (on buses 14, 8 and 12, based on the the magnitudes of the entries in the vectors $\sigma_n\bm{u}_{n}$, $n=1,\dots,3$) to transmit for the electrical structure-based approach, while continuing with 4 PMUs for the topology-based approach. The scheduling scheme is applied to both these cases and the resulting plots are shown in \figref{fig:Pd_time_elecless_topo}. As shown in the \figref{fig:Pd_time_elecless_topo}, the probability of detection for the electrical structure-based approach is in fact slightly better than the topology-based approach. It is important to note that, the above two modifications in the experimental setup are especially useful when there are stringent constraints ({\eg}, bandwidth limitation) for communications on the grid.

\begin{note}\label{note:fig5_6}
For \figref{fig:Pd_time_elec_topo} and \figref{fig:Pd_time_elecless_topo}, it should be noted that we have installed the optimum number of 7 PMUs for the electrical structure-based approach, so as to achieve complete network observability. However, we allowed only 4 PMUs to transmit their phasor data to the control center. Though the remaining 3 PMUs record phasor data across the lines on which they are installed, they do not transmit them to the control unit. So long as phasor data from the ``high priority'' buses ({\ie}, buses having high electrical influence on the remaining buses in the network) are obtained, the test statistic can be computed and the hypotheses test can be tested.
\end{note}

In summary, since the scheduling is based on the electrical connectedness between buses in the grid, PMUs placed on buses having a strong electrical influence with other buses are given a higher priority. Furthermore, as discussed in \secref{sec:introduction}, the electrical structure-based approach provides a stronger characterization of the electrical influence between network components compared to the topology-based approach. Therefore, the probability of detection of change in the susceptance parameter is higher when the PMUs are installed based on the electrical structure-based approach compared to the case when they are installed using the topology-based approach. However, both approaches outperform the case when phasor data from PMUs are transmitted without prior scheduling.

The scheduling scheme proposed in this paper assumes dedicated direct communications link between the individual PMUs and the WAMS control center. However, in practice the communications infrastructure for the power grid encompasses many complicated and interrelated operations \cite{Shahidehpour2003}. For instance, the integrated control center system (ICCS) provides user interfaces to view the power grid information and distributed computing environment to monitor and coordinate the security of transmission system. Then there is the paradigm of common information model (CIM) using which information sharing of power system applications can be achieved using a common markup language. Such communications overheads could be termed as ``constraints'', since they impede the rate of transfer of synchrophasor data from the PMUs to the WAMS server. However, as mentioned in \secref{sec:introduction}, since our scheduling policy is transmitter-centric, it is unaffected by the communications constraints. It should also be noted that integration of PMU schedulers into existing standards ({\eg}, C37.118.2-2011 \cite{2011}) could have a bearing on communication protocols, data types and formats for phasor data transmission on the grid. These aspects are relegated to future work.

\vspace{-0.1in}
\section{Conclusion}\label{sec:conclusion}
We proposed a time division multiplexing scheme for transmission of phasor data from the PMUs to a central server, where time was divided into frames and the optimal set of PMUs within a given time frame take turns to transmit to the control center. The proposed scheduling policy was governed by the measure of electrical connectedness between buses in the power grid. We presented the PMU placement problem from two different perspectives, namely, topology- and electrical structure-based approaches. For both these approaches, the scheduling scheme was coupled with a fault detection algorithm, which was posed as hypothesis testing problem. The performance of the fault detection, which was formulated to detect changes in the susceptance parameters of the network, was shown to improve due to scheduling transmissions from PMUs compared to transmitting phasor data in a random manner. Future work would involve scheduling algorithms with incomplete network observability and WAMS servers with multiuser reception/detection capability.

\vspace{-0.1in}
\appendices
\section{On the electrical structure of the grid}\label{app:electrical_structure}
The electric power grid has received considerable attention from the perspective of complex networks \cite{Dorfler2010}. In the following we briefly present this perspective, which promotes the electrical structure of the grid over its topological structure. Then, the binary connectivity matrix is derived using the resistance distance between buses in the network.

The study of the electrical structure of the grid was motivated by the following drawbacks suffered by its topological structure:
\begin{enumerate}
\item In \cite{Cotilla-Sanchez2012} (see Section I and references therein), it was reported that electric grids
in different geographical locations had different degree distributions leading to varied topological structures.
\item It was also pointed out that the same grid had different topological structures by carrying out different model-based analyses. This discrepancy was attributed to the weaker characterization of the electrical connections between network components as provided by the topological structure.
\item Related reports supporting this line of argument were found in \cite{Wu1995} - \nocite{Wu2005}\cite{Atay2006}, where it was shown that, for many classes of complex networks, characterizing the network structure using degree distribution alone was suboptimal and had implications on node synchronization and performance of the network.
\end{enumerate}

In the context of PMU placement, for the topology-based approach, the bus admittance matrix plays a central role in solving the placement problem. Though the admittance matrix characterizes the electrical behavior of the network, the sensitivity between power injections and nodal phase angle differences can be utilized to better characterize the electrical influence between network components. Towards this end, we derive the resistance distance matrix, which provides a strong characterization of the electrical influence between various network components.

Consider a network with $B$ buses, described by the conductance matrix $\bm{G}$. Let $V_j$ and $g_{ij}$ denote the voltage magnitude at bus $j$ and the conductance between buses $i$ and $j$, respectively. The current injection at bus $i$ is then given by
\begin{eqnarray}\label{eq:current_injection}
I_i = \sum_{j=1}^{B}g_{ij}V_j.
\end{eqnarray}
$\bm{G}$ acts as a Laplacian matrix to the network, provided there are no connections to the ground, {\ie}, if $\bm{G}$ has rank $B-1$. The singularity of $\bm{G}$ can be overcome by letting a bus $r$ have $V_r = 0$. The conductance matrix associated with the remaining $B-1$ buses is full-rank, and thus we have
\begin{eqnarray}\label{eq:nonreferencenodes}
\bm{V}_k = \bm{G}^{-1}_{kk}\bm{I}_k, k \neq r.
\end{eqnarray}
Let the diagonal elements of $\bm{G}^{-1}_{kk}$ be denoted $g^{-1}_{kk}$, $\forall k$, indicating the change in voltage due to current injection at bus $k$ which is grounded at bus $r$. The voltage difference between a pair of buses $(i,j)$, $i\neq j\neq r$, is computed as follows:
\begin{eqnarray}\label{eq:voltage_difference}
e(i,j) = g^{-1}_{ii} + g^{-1}_{jj} - g^{-1}_{ij} - g^{-1}_{ji},
\end{eqnarray}
indicating the change in voltage due to injection of $1$ Ampere of current at bus $i$ which is withdrawn at bus $j$. $e(i,j)$ is called the resistance distance between buses $i$ and $j$, and describes the sensitivity between current injections and voltage differences. In matrix form, letting $\boldsymbol{\Gamma} \triangleq \text{diag}(\bm{G}^{-1}_{kk})$, we have $\forall k \neq r$
\begin{eqnarray}
\bm{E}_{kk} &=& \boldsymbol{1}\boldsymbol{\Gamma}^{\mathrm{T}} + \boldsymbol{\Gamma}\boldsymbol{1}^{\mathrm{T}} - \bm{G}^{-1}_{kk} - \left[\bm{G}^{-1}_{kk}\right]^{\mathrm{T}},\label{eq:resistance_distance1}\\
\bm{E}_{rk} &=& \boldsymbol{\Gamma}^{\mathrm{T}},\label{eq:resistance_distance2}\\
\bm{E}_{kr} &=& \boldsymbol{\Gamma}.\label{eq:resistance_distance3}
\end{eqnarray}
The resistance distance matrix $\bm{E}$, thus defined, possesses the properties of a metric space \cite{Klein1993}.

To derive the sensitivities between power injections and phase angles, we start with the upper triangular part of the Jacobian matrix obtained from the power flow analysis, for the distance matrix to be real-valued:
\begin{eqnarray}\label{eq:upper_jacobian}
\Delta \bm{P} = \left[\frac{\partial P}{\partial \theta}\right]\Delta \theta + \left[\frac{\partial P}{\partial |V|}\right]\Delta |V|.
\end{eqnarray}
The matrix $\left[\frac{\partial P}{\partial \theta}\right]$ will be used to form the distance matrix, by assuming the voltages at the buses to be held constant, {\ie}, $\Delta|V|=0$. It was observed that $\left[\frac{\partial P}{\partial \theta}\right]$ possesses most of the properties of a Laplacian matrix. By letting $\bm{G} = \left[\frac{\partial P}{\partial \theta}\right]$, the resulting distance matrix $\bm{E}$ measures the incremental change in phase angle difference between two buses $i$ and $j$, $(\theta_i - \theta_j)$, given an incremental average power transaction between those buses, assuming the voltage magnitudes are held constant. It was proved in \cite[Appendix]{Cotilla-Sanchez2012} that $\bm{E}$, thus defined, satisfies the properties of a distance matrix, as long as all series branch reactance are nonnegative.

For a power grid with $B$ buses and $K$ branches, the distance matrix $\bm{E}$ translates into an undirected graph with $B(B-1)$ weighted branches. In order to compare the grid with an undirected network without weights, one has to retain the $B$ buses, but replace the $K$ branches with $K$ smallest entries in the upper or lower triangular part of $\bm{E}$. This results in a graph of size $\{B, K\}$ with edges representing electrical connectivity rather than direct physical connections. The adjacency matrix $\bm{C}$ of this graph is obtained by setting a threshold, $\lambda$, adjusted to produce exactly $K$ branches in the network:
\begin{eqnarray}\label{eq:adjcency_matrix}
\bm{C}:
\begin{cases}
\tilde{c}_{ij} = 1, ~\forall e(i,j) < \lambda,\\
\tilde{c}_{ij} = 0, ~\forall e(i,j) \geq \lambda.
\end{cases}
\end{eqnarray}
To obtain the threshold $\lambda$ present in \eqref{eq:adjcency_matrix}, we first consider the upper triangular part of the matrix $\bm{E}$, and sort the elements in descending (or ascending) order. We then pick a number of elements equal to the number of branches in the given power network. For instance, given the IEEE-14 bus network having 20 branches, we pick the top 20 sorted elements from the upper triangular part of the matrix $\bm{E}$.

\bibliographystyle{IEEEtran}
\bibliography{IEEEabrv,powersystems}

\end{document}